### 23.2 A 28nm 0.22μJ/Token Memory-Compute-Intensity-Aware CNN-Transformer Accelerator with Hybrid-Attention-Based Layer-Fusion and Cascaded Pruning for Semantic-Segmentation

Pingcheng Dong*[1,2], Yonghao Tan*[1,2], Xuejiao Liu[2], Peng Luo[2], Yu Liu[2], Luhong Liang[2], Yitong Zhou[2], Di Pang[2], Man-To Yung[2], Dong Zhang[1,2], Xijie Huang[1,2], Shih-Yang Liu[1,2], Yongkun Wu[1,2], Fengshi Tian[1,2], Chi-Ying Tsui[1,2], Fengbin Tu[1,2], Kwang-Ting Cheng[1,2]

[1]The Hong Kong University of Science and Technology, Hong Kong, China
[2]AI Chip Center for Emerging Smart System, Hong Kong, China

*Equally Credited Authors (ECAs)

Recently, hybrid models integrating a CNN and a Transformer (ConvFormer), shown in Fig. 23.2.1, have achieved significant advancements in semantic segmentation tasks [1-4], which are critical for autonomous driving and embodied intelligence. The CNN enhances the multi-scale feature extraction ability of the Transformer to achieve pixel-level classification, but the large token length (TL) demand of semantic segmentation (>16K TL) incurs significant computation and memory overheads. Prior NN accelerators [5-12] demonstrate that sparse computing and pruning can effectively reduce computation and weight storage, but most of them focus on pure CNN or Transformer models in simpler vision or language-processing tasks (1-4K TL). Moreover, the performance bottlenecks of ConvFormers stem from their memory-intensive Backbone and compute-intensive Segmentation Head (Seg. Head), raising three challenges for hardware acceleration: 1) Conventional sparse attention [5-9] fails to buffer the attention feature map (Fmap) on-chip when the TL exceeds 16K, even at 90% sparsity, resulting in massive external memory access (EMA). 2) While Layer-Fusion (LF) [13-18] is a common technique to reduce Fmap EMA, it is infeasible to buffer key (K), value (V), and convolution weights on-chip simultaneously. Moreover, different fused attention-convolution layers may cover various vanilla attention (VA) tiles, leading to enormous redundant KV and weight EMA. 3) In the Seg. Head, the Fmap sparsity is extremely low, thereby limiting the effectiveness of conventional zero-skipping strategies [11,12] designed to reduce computational work.

To tackle these challenges, we propose a ConvFormer accelerator with three key features: 1) We propose a Hybrid Attention Processing Unit (HAPU) that utilizes memory-efficient linear attention (LA) [19-23] for most query (Q) tiles in Backbone, reducing Fmap storage from $O(N^2)$ to $O(C^2)$ by reordering the computation from $QK^T$-first to $K^TV$-first, while a few Q tiles perform VA to ensure accuracy by their global receptive field. Here, N is the TL and C is the channel dimension with C<<N. This reordering allows the HAPU to buffer tiny $K^TV$ Fmaps entirely on-chip, saving 60.2-78.6% EMA. 2) We develop an LF Scheduler (LFS) with KV-weight reuse to mitigate redundant EMA overhead of LF in Backbone. The LFS first reuses on-chip buffered KV to compute all VA tiles, then replaces the KV with off-chip convolution weights. Afterward, convolution layers are sequentially fused with each VA output tile and LA input tile, reusing both $K^TV$ and convolution weights. This approach significantly alleviates redundant EMA, reducing overall EMA by 86.8-96.2%. 3) A Cascaded Fmap Pruner (CFMP) is designed to decompose each convolution of Seg. Head into two sub-convolutions: the first injects redundancy by expanding the intermediate Fmap, which is further pruned using a pre-trained mask, while the second restores density using the same mask, reducing 91.10% of computation in the Seg. Head. Notably, while this work focuses on accelerating ConvFormer for semantic segmentation tasks, the proposed solutions are not limited to ConvFormer and can be also applied to pure CNN or Transformer models.

Figure 23.2.2 shows the overall architecture of the proposed ConvFormer accelerator. It consists of a SIMD core, a top controller, a PLL, 2 HAPUs, a LFS, a CFMP, a 64KB ISA buffer, a global buffer (GB) including a 2MB left matrix buffer (LMB), a 1MB right matrix buffer (RMB) and an LMB-RMB Router ($LR^2$). In the Backbone stage, the HAPUs prioritize $K^T$, V, and $K^TV$ generation, where $K^T$ is further routed from LMB to RMB via $LR^2$. Then, LFS clusters the VA and LA tiles in the attention cluster unit (ACU), scheduling HAPUs to reuse KV for parallel VA tile processing. Once completed, the LFS replaces the KV with subsequent convolution weights, directing HAPUs to perform fused convolution on each VA output tile. Then, the remaining convolution layers could reuse these weights to fuse with their associated LA input tiles. In the Seg. Head stage, the top controller configures CFMP using pre-trained sparsity masks. The feature map sparsifier (FMS) decodes the RMB IDs of unpruned column tiles, which are sent to HAPUs for sparse convolution. Then, the density recovery unit (DRU) converts column tile IDs in FMS to row tile IDs, guiding HAPUs to recover the density of sparse Fmap via row-wise accumulation.

Figure 23.2.3 illustrates HAPU that leverages a hybrid attention mechanism for EMA reduction. This hybrid attention allows most Q tiles to employ LA, replacing the exponential function of VA with separable kernel functions, such as identity, ReLU, etc. The hybridization pattern is learned during training and exhibits a layer-wise distribution. However, since $K^T$ serves as a right matrix for $QK^T$ in VA and left matrix for $K^TV$ in LA, it incurs storage conflicts that require $K^T$ transfers through DDR. To address this issue, the right matrix prioritized initializer (RMPI) first computes $K^T$, V, and $K^TV$, then decodes the $LR^2$ offset to route $K^T$ from LMB to RMB on-chip via $LR^2$. In addition, certain layers may retain a high proportion of VA, leading to large VA tiles that could still cause EMA issues. To mitigate this, the Attention Tiling Manager (ATM) identifies the VA tile size and speculates potential LMB overflows. If overflows are detected, the ATM subdivides each Q tile into smaller segments, which are processed sequentially and combined together. Compared to a layer-wise architecture, the HAPU achieves a reduction in EMA and energy consumption by 22.05× and 8.93×, respectively, for an attention layer with a 64K TL and 32 channel dimensions.

Figure 23.2.4 depicts the LFS that consists of an ACU, a KV-reused vanilla attention-convolution fuser (VACF), and a weight-reused linear attention-convolution fuser (LACF). Owing to the significant reduction in VA proportion by HAPU, all VA tiles can be computed in parallel for most layers. The ACU initially re-orders the VA and LA tiles into two fusion groups, integrating them with the same fused convolution (FC). Subsequently, the VACF decodes the LMB ID for each VA tile to generate Q and reuses the KV prepared by RMPI for parallel VA execution. The VACF then fetches off-chip FC weights to replace KV, and sequentially schedules each VA output tile for its corresponding FC. Since the $K^TV$ is prepared in advance by RMPI and remains on-chip due to its small size, and the FC weights are already buffered on-chip by VACF, the LACF can reuse them to perform LF from each LA input tile. However, the convolution cannot fuse with boundary tiles between VA and LA in VACF because the LA output tiles are not yet ready. While the previous slice-based LF method [16] can handle this with overriding, it incurs extra storage and computation overhead for each VA tile. To address this, we propose a non-overlapped LF processing scheme wherein the FC is broken into several non-overlapped FCs. The subsequent attention layer recovers the broken receptive field by its inherent long-range dependency. The boundary fusion issue is then resolved by zero-padding the unavailable boundary tiles, resulting in 50% GB usage and a 20% reduction in operations, with <0.5% accuracy drop. Moreover, EMA and energy consumption are reduced by 3.91× and 1.45×, respectively, for a ConvFormer sub-block with a 64K TL and a 50% VA ratio.

Figure 23.2.5 introduces the workflow and architecture of CFMP, which contains an FMS and a DRU. The CFMP decomposes the convolution weights into two cascaded components $W_0$ and $W_1$, injecting redundancy by enlarging the intermediate Fmap Z, which is then pruned using pre-trained tiled masks. Since Z is the only sparse Fmap, with the input and output Fmaps remaining dense, CFMP can be generalized to support sparse VA by substituting the input Fmap, $W_0$, $W_1$, and the output Fmap in convolution with Q, $K^T$, V, and the output Fmap in VA. The FMS begins by decoding the mask and splitting it into multiple parts for pipeline processing. Tiled column (TC) offsets are generated by flattening each part and extracting the positions of non-zero values. Once all valid indices are decoded, the FMS halts the current mask decoding and fetches the next one, allowing for an early stop. The TC offsets are combined with the base IDs from the LMB/RMB and sent to the HAPU, where the unpruned TCs are fetched, and the sparse Z is stored in a dense format. Then, the DRU needs to obtain unpruned Tiled Rows (TRs) corresponding to each Z tile and its associated mask to recover the Fmap density. However, the physical storage scheme of the RMB maps each TC across different SRAM banks consecutively, resulting in interleaved storage of various TR slices within the same bank from a row-wise perspective. To address this, the DRU converts TC offsets into TR form and recovers density by accumulating the multiplication results from different Z tiles and TR slices. Compared to zero-skipping approaches [11,12], the CFMP improves sparsity by 6× in the Seg. Head and reduces energy consumption by 2.03× when pruning both VA and convolution.

Figure 23.2.6 presents the measurement results for the ConvFormer accelerator fabricated using a 28nm process. The chip works at 200-625MHz with a supply voltage of 0.65-1.0V. The peak energy efficiency is 52.90TOPS/W at 0.65V and 200MHz. Experiments are conducted on three ConvFormer models, SegFormer-B0 [1], PVTv1-Ti [2], and PVTv2-B0 [3], with the Cityscapes dataset. The memory-intensity-aware HAPU and LFS and compute-intensive-aware CFMP obtain 4.66-to-7.71× speedup and 4.39-to-7.10× energy savings compared to the baseline, with negligible accuracy loss. Furthermore, we include a DDR3 interface similar to [10] and assume all prior state-of-the-art accelerators [5-8, 10] work at peak energy efficiency with their reported technical configurations, such as pruning ratio, sparse attention patterns, etc., to evaluate system energy consumption fairly. The results indicate that our chip consumes 0.22μJ/token for SegFormer-B0, achieving 3.86-to-10.91× system-level energy reduction. The die photo, voltage-frequency scaling curves, and specification table are shown in Fig. 23.2.7.

*Acknowledgement:*
This research was supported by ACCESS – AI Chip Center for Emerging Smart Systems, sponsored by InnoHK funding, Hong Kong SAR. The corresponding authors of this paper are Kwang-Ting Cheng (timcheng@ust.hk) and Fengbin Tu (fengbintu@ust.hk).

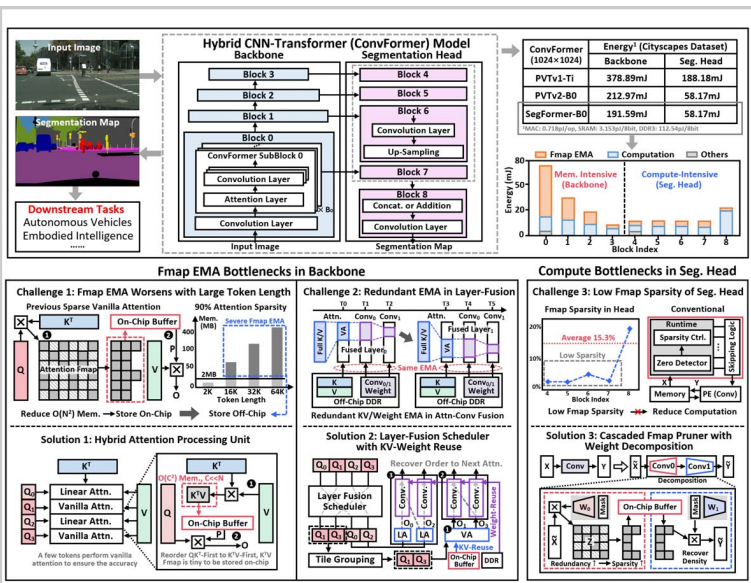

Figure 23.2.1: Hybrid CNN-Transformer (ConvFormer) models raise three challenges for ConvFormer accelerator design.

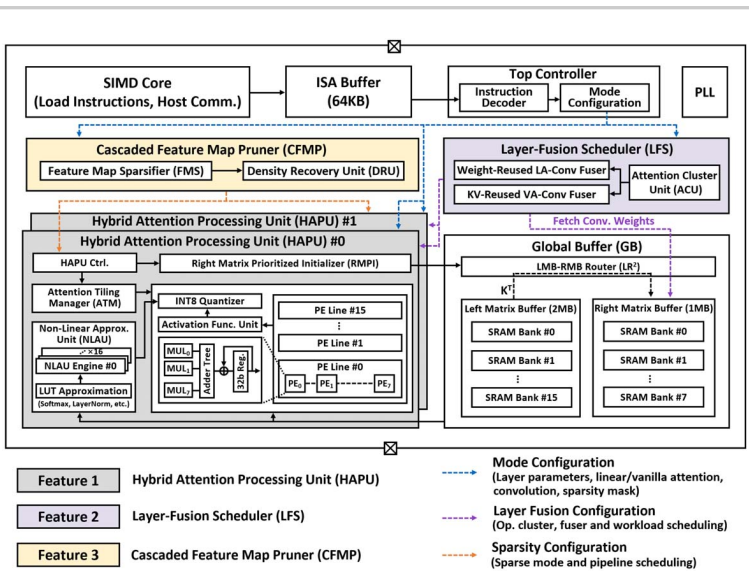

Figure 23.2.2: Overall architecture of the proposed ConvFormer accelerator and three main features.

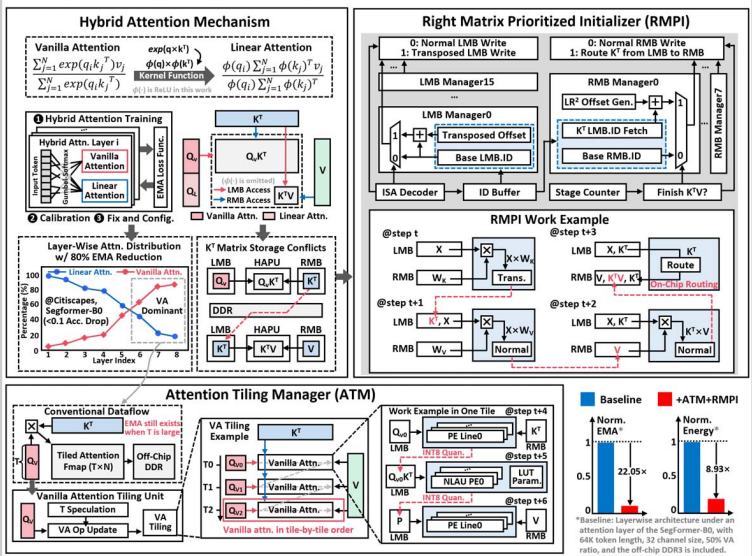

Figure 23.2.3: Attention Tiling Manager (ATM) and Right Matrix Prioritized Initializer (RMPI) in Hybrid Attention Processing Unit (HAPU) for attention EMA reduction.

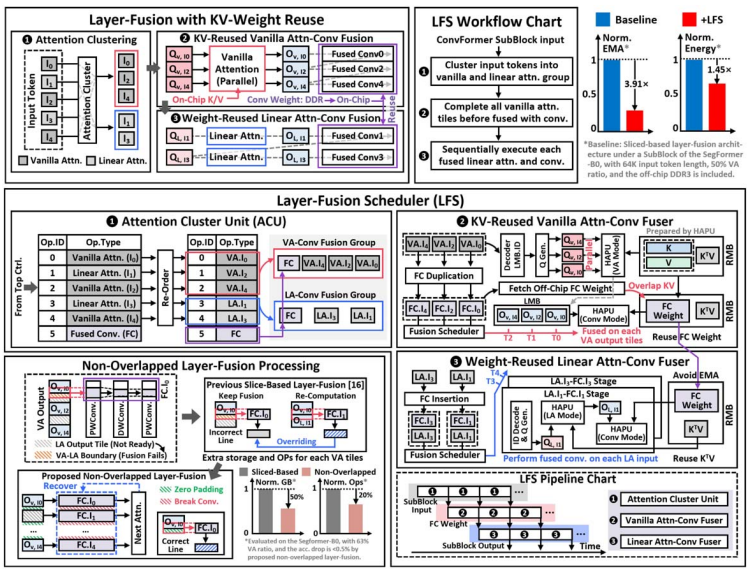

Figure 23.2.4: Data-Reuse Layer-Fusion Scheduler (DR-LFS) for redundant EMA-eliminated attention-convolution layer-fusion.

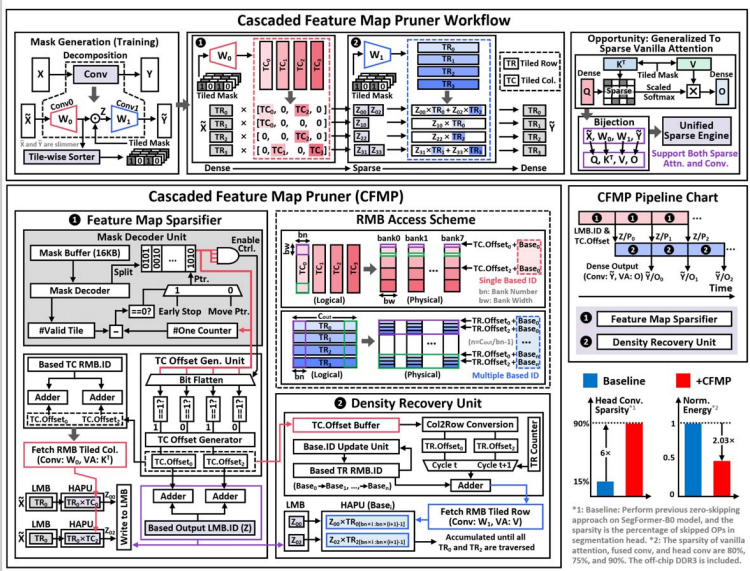

Figure 23.2.5: Cascaded Feature-Map Pruner (CFMP) for sparsity optimization.

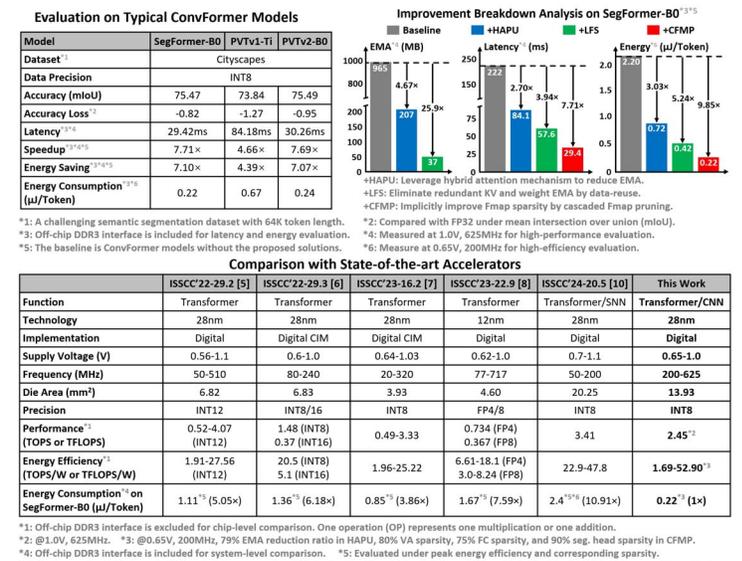

Figure 23.2.6: Measurement results and performance comparison table.

23

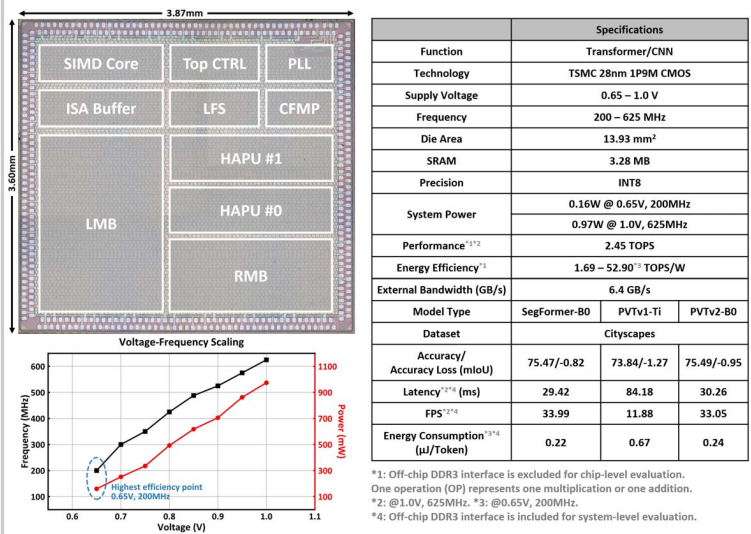

Figure 23.2.7: Die photo, voltage-frequency scaling curve, and specification table.